\documentstyle[11pt,newpasp,psfig]{article}
\def\mb#1{\setbox0=\hbox{$#1$}\kern-.025em\copy0\kern-\wd0
\kern-0.05em\copy0\kern-\wd0\kern-.025em\raise.0233em\box0}

\def\edcomment#1{\iffalse\marginpar{\raggedright\sl#1\/}\else\relax\fi}
\reversemarginpar

\def\mb#1{\setbox0=\hbox{$#1$}\kern-.025em\copy0\kern-\wd0
\kern-0.05em\copy0\kern-\wd0\kern-.025em\raise.0233em\box0}
\def\gsim{\lower 2pt \hbox{$\, \buildrel {\scriptstyle >}\over
{\scriptstyle \sim}\,$}}
\def\lsim{\lower 2pt \hbox{$\, \buildrel {\scriptstyle <}\over
{\scriptstyle \sim}\,$}}

\begin{document}
\title{Pulsar Spin-Down Induced Phenomena: Heating; Magnetic Field
Evolution; Glitches; Pulse-Period Modulations}
\author{M. Ruderman}
\affil{Physics Department \& Columbia Astrophysics Laboratory, Columbia
University, 550 W. 120th Street, New York, NY  10027, USA}

\begin{abstract}Modeling the dynamics of the quantum fluids within a
spin\-ning-down neutron star gives a description consistent with observed
pulsar magnetic field evolution and spin-period ``glitches."  The long-standing problem of
large predicted excesses in spin-down sustained pulsar heating from such models now seems
resolvable. However, the origin of some  pulsar spin-period and pulse-shape modulations which
have been interpreted as manifestations of very long period ($\sim$ year) stellar
precession is a crucial challenge to canonical neutron star models.

\end{abstract}

\section{Introduction}

In a cool core below the crust of a spinning neutron star (NS) superconducting
protons coexist with the more abundant superfluid neutrons
to form a giant atomic nucleus which includes a neutralizing sea of relativistic
degenerate electrons. The neutrons rotate with a spin-period $P$ (sec) only by forming a
nearly uniform array of corotating quantized vortex lines parallel to the spin axis,
with an area density
$n_v 
\sim10^4{\rm
cm}^{-2}P^{-1}$.  The array contracts (expands) when the star
spins-up (down).
For stellar core neutron spin-up or spin-down, a vortex a distance
$r_\perp$ from the spin axis generally moves with a velocity ${\mb{v}}_v
= {\mb{r}}_\perp \dot P/2P$ until $r_\perp$ reaches the core neutron superfluid
radius $(R)$. 
Any stellar magnetic field passing below the stellar crust
must, in order to penetrate through the superconducting protons, become bunched
into a very dense array of quantized flux tubes ($n_\Phi \sim 5 \times
10^{18}B_{12}{\rm cm}^{-2}$ with $B$ the local average magnetic field). 
Each tube carries a flux $2 \times 10^{-7}{\rm G cm}^2$ and magnetic field $B_c
\sim 3\times 10^{15}{\rm G}$.  [This assumes a Type II proton superconductor
below the crust, the result of essentially all calculations.  If it were Type I,
details would change but not the conclusions below.] The initial magnetic field
within the core of a neutron star is expected to have both toroidal and very
non-uniform poloidal components.  The web of flux tubes formed after the transition
to superconductivity is then much more complicated and irregular than the neutron
vortex array as well as of order $10^{14}$ more dense.

Enough is understood about the dynamics of the components of a canonical NS to allow
what should be a reliable, reasonably quantitative, description of what happens
within a spinning magnetized NS as it ages and spins-down, and also in those rarer
cases where it is spun-up by a companion.  Because of the velocity dependence of the
short range nuclear force between neutrons and protons there is a strong interaction
between the neutron superfluid's vortex lines and the proton superconductor's flux
tubes if they come closer to each other than about $10^{-11}$cm. Consequently,
when $\dot P \neq 0$ flux tubes will be pushed
(or pulled) by the moving neutron vortices [Sauls 1989, Srinivasan et al 1990,
Ruderman 1991, Ding, Cheng, \& Chau 1993, Ruderman, Zhu, \& Chen 1998, Jahan-Miri
2000, Konenkov \& Geppert 2001]. A realistic flux-tube array will be forced to move
along with a changing neutron-superfluid vortex array which threads it as long as the
force at a vortex-line flux-tube juncture does not grow so large that vortex lines
cut through flux tubes. This is the basis for predicting an evolution of pulsar
magnetic fields during spin-down or spin-up which seems to agree well with pulsar
observaitons.  However, it also leads directly to two crucial problems which could
be ``show stoppers'' for modeling neutron star interiors if unresolved.  We discuss
the first of these next.

\section{Heat Generation Inside Young Spinning-Down Pulsars}

Just as in the case of a continuously disturbed magnetic field in a classical
conducting fluid forced motion of a large flux tube array {\it through} the
electron-proton sea in which it is imbedded could occur only if there is local ohmic
dissipation.  The main contribution to such dissipation is from the random part of
electron scattering on the flux tube lattice [Ruderman et al. 1998]. If there is no
flux-tube cutting so that  all flux-tubes are pushed through a core's 
electron-proton sea, the heat production rate from it ($\dot Q$) has been calculated
to be about $10^{35}$erg~s$^{-1}$ within the $P=0.9$~s Vela pulsar.
A similar heating is predicted if there is flux tube cut-through in a  Vela-like
pulsar. But soft X-ray observations of the $10^4$yr old Vela  give a bound of $\dot Q
< 10^{33}{\rm erg\ s}^{-1}$ [\"Ogelman, Finley, \& Zimmerman 1993]. 
Therefore, 
a crucial question for  all spin-down
models of pulsars with strong interior magnetic fields is
 how moving core vortex lines move with, or through,
the extraordinarily dense flux tube array in which they are imbedded without an
unacceptably large $\dot Q$. (Special {\it ad hoc} magnetic field configurations
which could greatly diminish $\dot Q$ appear implausible and also inconsistent with 
classical magnetohydrodynamical stability much earlier before cooling below
the transitions to neutron superfluidity (vortices) and proton superconductivity
(flux-tubes).)
By far the most promising answer seems to be the huge
drag reduction because of flux-tube clumping from the expected instability 
when  flux-tubes are pushed by the somewhat flexible neutron superfluid
vortex-lines.
Vortex lines move  a  
layer of flux-tubes  as indicated in Fig. 1a.  When the
moving flux tubes, which repel each other
strongly only at  distances $\lsim 10^{-11}$cm,
form a ``thick'' blanket (Fig. 1b), their movement generates 
electron  currents through the flux tube
array which must be dissipated for the blanket to move 
through the
$e-p$ fluid.
\begin{figure}
\psfig{file=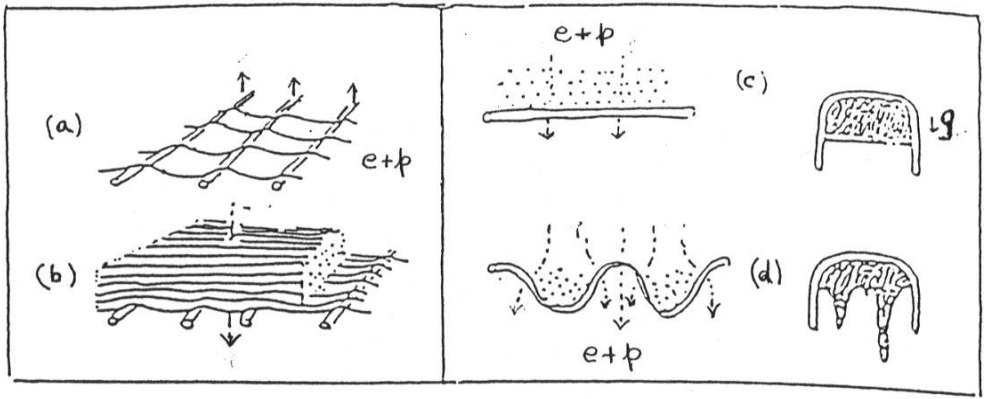,width=\linewidth}
\caption{Flux-tube (lines) clumping by moving vortex lines
(tubes).}
\end{figure}
Figure 1c is a side view of the vortex
flux tube system in a reference frame where the vortex lines are at rest.  The balance of
forces between the force density from electron flow through flux tubes, the
 push of the vortex lines (the ``Magnus force'' which pushes them
outward as the star spins-down), and the tension in the (bent) vortex lines, is
exactly analogous to the pull of gravity $(\mb{g})$ in an upside down glass of
water, the upward push of the atmosphere, and the surface tension at the water-air
interface.  Both systems are unstable in the same (Rayleigh-Taylor) way, and the
interface instability quickly becomes large and non-linear.  Field-free spaces develop
between clumps of flux tubes, through which unmagnetized $e-p$ fluid flows (Fig.
1d). Moving $e-p$ fluid in the vortex rest frame  no longer passes through
the flux-tube bundles.  The resistance to the bunched flux tube flow  in the $e-p$
rest frame is then  only the viscosity of the unmagnetized degenerate electron sea
around the bundles or, if the effective $e-e$ scattering mean free path of these
electrons exceeds the radius of the flux tube bundles, the scattering of electrons
by them.  Preliminary numerical estimates for the scale of the instabilites and the
drag on the moving bunches does indeed give a $\dot Q$ very much smaller than the
upper bound allowed by the soft X-ray observations. Thus, except for possible
constraints from the anchoring of the core's flux-tubes by the surrounding
conducting crust, the expanding (contracting) neutron superfluid vortex array in the
core of a spinning-down (up) pulsar should relatively easily induce a similar
expansion (contraction) of the core's flux-tube array.

\section{Surface Magnetic Fields of Spinning-Down Neutron Stars}

The core of a cool neutron star is surrounded by a highly conducting $10^5$cm
thick crust which anchors any magnetic field through it until either the crust
yield strength is exceeded by ${\mb{j}} \times {\mb{B}}$ forces in it 
and the most strongly magnetized parts of the crust follow the core flux
tube movement below
or
eddy current decay allows crust flux movement.  During various
epochs in the life of a neutron star, this insures that its surface field will
ultimately reflect that of the high flux density regions anchored at a crust's
base.  All of this leads to the evolutionary track shown in Fig. 2 for the surface
dipole magnetic field of a pulsar as its period ($P$) changes. The ``observed" NS's
surface dipole field $B$ is that inferred for the dipole moment ${\mb{\mu}}$ from
$I\dot\Omega \simeq - \mu^2\Omega^3 c^{-3}$. ($I$ is the NS moment of inertia.) 
  The dipole evolutionary
curve of Fig. 2 has the following segments. 

\begin{figure}
\centerline{\psfig{file=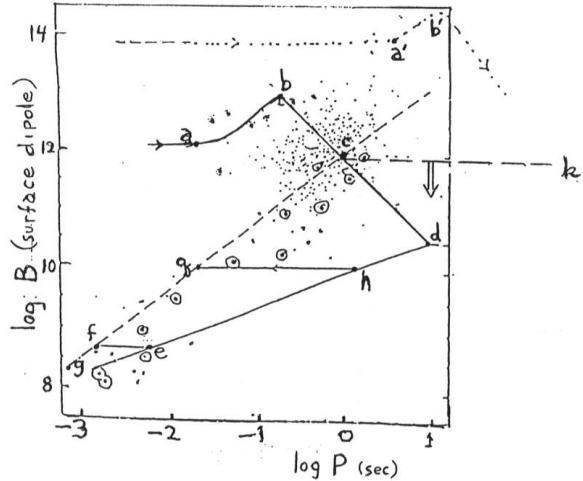,width=3truein}}
\caption{Model evolution of magnetic dipole field
and observed values inferred from spin-down rates and periods of
radio-pulsars. Starlike designations correspond to very young
radio-pulsars in SNR's.  Circled points are pulsars in binaries, usually
 spun-up candidates.  The dashed line is the accretion-determined
``spin-up line." The dotted line is for magnetar evolution.}
\end{figure}

{\it Pulsar birth until $a$: $B$ is constant.} 
Core-vortex motion-induced field changes do not begin until the NS cools enough
for both flux-tube and vortex-line formation. 
 For
$B \sim 3\cdot 10^{12}$G, $P$ will have grown to $P_0 \sim 20$ms before the required
cooling is achieved no matter how much smaller the initial $P$ may have been.
(For a ``magnetar" with $B\sim 10^{15}$G, $P_0\sim 7$s.)

$a\rightarrow b:$ Here $r_\perp \propto P^{1/2}$ until $r_\perp$ reaches
the stellar radius $R$. 
The predicted evolution  of $\vert {\mb{\mu}} \times \hat{\mb{\Omega}}\vert \equiv
\mu_\perp$ is then  particularly simple. Models which attribute
spin-down mainly to the Maxwell torque,  $I\dot\Omega \sim
\mu_\perp^2\Omega^3c^{-3}$ have a
``spin-down index" $n\equiv -\Omega\ddot\Omega\dot\Omega^{-2} =3 -
{2\dot\mu_\perp\Omega\over \mu_\perp\dot\Omega}$. Because
 $\mu_\perp$ grows larger
with increasing $P$, 
$n$
grows from 2 to 3  until much of the surface flux is pushed out to $r_\perp
\sim R$.  These predicted $n$ values compare reasonably with those observed in  
young high spin 
radiopulsars with  measured $n$ (Crab, $n=2.5$ [Lyne, Pritchard, \& Smith 1998]; PSR
1509-58,
$n=2.8$ [Kaspi et al. 1994]; PSR 0540, $n=1.8$ [Zhang et al 2001]; PSR J1119,
$n=2.9$ [Camilo et al. 2000]). 

$b\rightarrow c:$ As flux tubes are pushed out of the core
by the expanding vortex array, surface North and South poles are, at last, able to
respond to the great pull of the large flux pushed into the lower crust and move to
give surface field reconnection.  Thereafter
 $\vert \mu_\perp \vert$ decreases roughly as $P^{-1}$ 
$({\rm i.e.,} n=5)$, but the $n$ at a particular $P$ for any one pulsar cannot be predicted
without a detailed {\it a priori} knowledge of its magnetic field structure\footnote{The
Vela pulsar's suggested $n\sim 1.4$ [Lyne et al. 1996] is not incompatible with this model. 
Depending upon the original distribution of the core flux-tubes pushed to the crust-core
interface by Vela's expanding vortex array, the movement of a  North-South polar cap pair
which ultimately reconnects may be either along a shortest connecting path (decreasing dipole
and $n>3$) or along a maximum length great circle path (first an increasing dipole so that
$n<3$, followed later by a decreasing one with $n>5$ until the initial polar caps finally
overlap). Alternatively the surface field may begin with many North and South polar caps. 
The net dipole moment is then the resultant of several dipoles. One of these could become
smaller but the resultant increase.  Only after relatively long evolution is $\langle n
\rangle \sim 5$ realized. (Vela's presently increasing dipole moment may occur mainly in
sudden crust cracking events (``glitches'', cf \S4)).}.
The
absence of many observed canonical pulsars with $P>P$(Vela) $=0.9$s
but $B\geq B$(Vela) $\sim 3\cdot 10^{12}$G,  compared to
a plethora of those with lower $B$,  gives considerable support to the model
prediction of a dipole
$B$ which now no longer grows with increasing $P$ but  decreases as shown in
Fig. 2. Especially supportive is the reported factor of two difference between
observed kinematic and spin-down pulsar ages, 
which corresponds exactly to the predicted for $n=5$ [Cordes 2000].

$c\rightarrow k$ {\it and} $c\rightarrow d:$ The point $c$ is about where the
maximum expected magnetic stress in the crust no longer exceeds the crust's
yield strength.  The evolution of surface $B$ from further spin-down beyond
$c$ depends on time scales.  The core surface  $B$ should
follow the trajectory $c) \rightarrow d)$.  The crust surface field  would now follow it only
when 
spin-down is slow enough for crustal eddy current dissipation or
plastic creep $(P/\dot P > {\rm several} \times 10^6$yrs). On shorter time scales the crust
surface field remains frozen at the value it had at the point $c$ and  the
$c)\leftrightarrow k)$ segment could be followed during spin-down and spin-up of
X-ray pulsars in accreting binaries.  When these neutron stars become sufficiently
old, however, their surface dipoles would drop to the $c\leftrightarrow d$ segment.

$d\rightarrow h\rightarrow q$, $d\rightarrow e\rightarrow f$, $d
\rightarrow e \rightarrow g$:
Here a dead  radiopulsar  in a low mass X-ray binary (LMXB)
is assumed to be spun-up by accretion from its companion
so that magnetic flux is now squeezed inward toward the spin axis. 	
The superfluid vortex
velocity within a neutron star being spun-up to a millisecond period in an LMXB,
($\sim
10^{-9}\ {\rm cm\ s}^{-1})$ is	 so small that  a 
core flux tube motion which follows it seems inescapable
even if there were no flux-tube bunching.
None of the expectations for magnetic field evolution of NS's in such LMXBs 
are in conflict with observations. 
This includes a remarkably large fraction of
apparently orthogonal and also of nearly aligned rotators among the
disk population's fastest spinning
millisecond pulsars (MSP's) [Chen, Ruderman, \& Zhu 1998, Jayawardhana \& Grindlay
1996].  Especially supportive is  the magnetic field structure of PSR 1937+21
implied by its observed polarization
properties and their radio frequency dependence [Chen \& Ruderman 1993].  It is just
that expected for the fastest spun-up MSPs ($P=1.6$ms).  Recycled case
($d\rightarrow b$) pulsars should reach the Fig. 2 spin-up line with a much more
nearly canonical dipole field strength than that of most MSP's.  At least one such
radiopulsar in a binary seems to have been identified [Van den heuvel \&
Bitznaraki 1995].  Just below the spin-up line near it is also where most candidates
for nearly aligned pulsars (i.e. radiopulsars with anomalously broad pulse-widths)
are found, as expected from this evolutionary model). Very strong support for the
``squeezed flux''
MSP  model also comes from the thermal part of the X-ray emission expected from the
polar cap of the nearly aligned MSP PSR 0437. It has  been reported at this meeting [Tr\"umper
2001] to be from a surface hot spot about
$10^{-2}$ the size of a canonical (i.e., central dipole or uniformly magnetized NS) 
polar cap with the PSR 0437 spin-period. This is just what has been predicted for
the ``squeezed flux" spin-up model  for that pulsar [Chen et al. 1998].

\section{Glitches in Canonical Pulsar Spin-Periods}

 The expansion of a core
vortex array in a spinning-down pulsar would overstress the crust in two ways. 
First, outward moving vortices pull on the crust through the crust anchored flux-tubes which
they pull with them.  The crust could ``crack" from the resulting overstrain in both the
$a\rightarrow b$ and the $b\rightarrow c$ segments of Fig. 2.  Each maximum crust
displacement would be expected [Ruderman 1991, et al. 1998] to give a
$\Delta
\mu_\perp/\mu_\perp \sim +10^{-3\pm 1}(3-n)^{-1}$.  This is consistent in sign
and magnitude with the larger observed unhealed jumps $\Delta\dot P/\dot P \sim 5
\times 10^{-4}$ in the  family of weak Crab pulsar period glitches [Lyne, Smith, \&
Pritchard 1992].  These are by far the largest observed non-transient fractional changes in
any pulsar parameters after a
 glitch. (It is not yet
known if similar jumps occur in  larger glitches of other pulsars.)  Such crust
movement can cause small jumps in the crust spin-rate in two ways: (1) Crust-lattice pinned
vortices may be shaken loose. While this may no longer seem large enough to give the giant
Vela-like glitches [Jones 1998] it may still suffice for the much smaller Crab-like ones. (2) 
A new but necessary contribution is from the
 reduction
in pull-back on the outwardly moving {\it core} vortex lines by the moved crust's pinning of
core flux tubes.  The core vortices must adjust to it by an (non-instantaneous)
outward displacement.  This alone has been crudely estimated to give a crust glitch
frequency jump with a magnitude similar to those of the Crab ones if the
proposed reduction of $\dot Q$ by reduced flux tube drag  is valid.
A second family, the {\it giant} Vela-like glitches would come from a different kind
of sudden core vortex movement.  In the absence of a very dense
flux tube environment, outward moving core vortices would smoothly shorten and
then disappear as they reached the core's equator (Fig. 3. right hemisphere). 
However, the highly conducting crust strongly resists entry by the flux array which
moves with these vortices (Fig. 3 left hemisphere).  This pile-up of pushed flux
tubes in an equatorial annulus also prevents vortex line expulsion 
into the lower crust and disappearance there
until either vortex-line flux-tube cut-through events or a sudden ``cracking" of
overstressed crust  which allows field adjustments.  Until either begins the
superfluid in the annulus rotates with  period $P_0 (\ll P)$ since none of the
vortices which formed when the entire core had the period
$P_0$ have yet been able to leave the core.  Giant Vela-like glitches are
proposed as the events which
allow a sudden reduction of this annulus of excess angular momentum by
outward movement of these trapped vortices
when the crust finally yields  to the magnetic stress at its base. These
would be expected along the $b \rightarrow c$ segment in Fig. 2, where reconnection
begins $(\bar n\sim 5)$, but before point $c$ is reached where crust strength
prevents further cracking.  This is indeed the segment on which giant glitches
have so far been observed.  The time scale between such glitches should be
related to the cracking displacement $(d)$.  If this is about the same as that 
in the larger Crab glitches, $d\sim \left({\Delta \mu_\perp/\mu_\perp}\right) R\sim
3\cdot 10^{-4}R$, there would then be $\sim R/d \sim 3\cdot 10^3$  giant
glitches during Vela's spin-down age or about one each three years.  This is
encouragingly close to that observed but details must be worked out. 

\begin{figure}
\centerline{\psfig{file=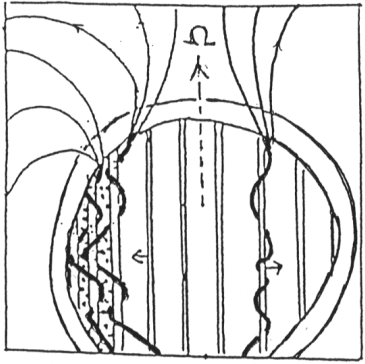,width=2truein}}
\caption{RHS. Vortices move outward and disappear smoothly
at the core equator.  LHS: Vortices squeeze flux tube array 
against the highly conducting crust which keeps
them from reaching the equator. In that outer annulus neutron superfluid retains its
initial period ($P_0$).}
\end{figure}

This model for the two families of pulsar spin-period glitches attirbutes both mainly to
phenomena involving spin-down of {\it core} neutron superfluid.  It differs from the
presently most widely applied 
models for  radiopular spin-up ``glitches" which
assume glitches to be caused by  a sudden reduction in rotation speed of the crust's neutron
superfluid filling the space between the crustal lattice nuclei.  That crust superfluid
is assumed to be prevented from spinning-down with the rest of the star because of
 pinning of its vortices to those nuclei  [Alpar et al 1984, 1993].
A glitch is supposed to be caused by a sudden collective unpinning. 
However, there appear to be significant unresolved problems with this kind of glitch model.
 a) The most
compelling is from  work of Jones [1998], who  calculated the propagation of 
vortex unpinning  in a crust.  He finds that this occurs typically two (or
more) orders of magnitude too easily to allow enough pinning 
for unpinning events
to give the observed ``giant glitches" in the
Vela pulsar spin rate ($\Delta P/P \sim - {\rm several\ } \times 10^{-6}$ at
intervals
$\tau_g
\sim 3$yr).  He  concludes: ``glitches do not originate in the
crust" (cf, however, Link and Cutler [2001]). b) Glitches are observed in two rather
separate groups: giant Vela-like glitches (upper arrow in Fig. 4) and a separate family with
glitches about
$10^{-2}$ as large [lower arrow in Fig. 4].  Individual pulsars 
on the $\overline{bc}$ segment of Fig. 2.
may have both
kinds.  Perhaps even more significant, giant Vela glitches seem to
begin with much larger initial spin-frequency jumps which quickly relax [McCulloch et al.
1990, Lundgren 1995] to those of Fig. 4, while the only Crab glitch observed from its
beginning (only
$10^{-2}$ as large as the Vela ones) shows quite the opposite initial behavior.
This dichotomy has not been easily explained in crustal vortex
unpinning models. By comparison, 
the existence of two separate glitch
families with  about the observed $\Delta P/P$ and  early time
evolutionary differences are  consequences of the magnetic field
evolution model of \S 3 if we add to it the common assumption that
crustal stresses which grow to exceed significantly the crust's yield strength
are relieved by  sudden (less than $10^2$s) crust  displacements (``cracks" or ``starquakes").

\begin{figure}
\centerline{\psfig{file=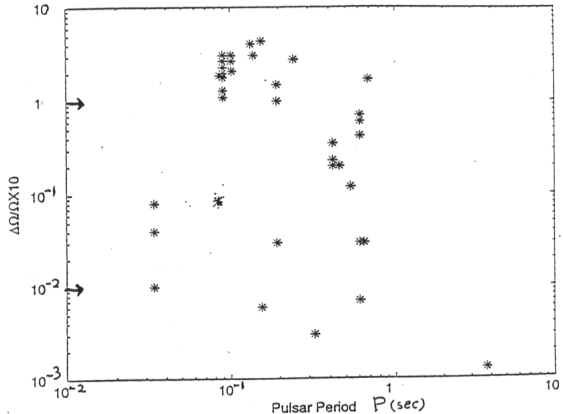,width=3truein}}
\caption{Pulsar glitch magnitudes $(\Delta\Omega / \Omega \times 10^6)$ in
various spinning-down pulsars. Data from Lyne [1995].}
\end{figure}

Up to this point the above description of pulsar evolution --- based upon expected vortex,
flux-tube, and crust dynamics --- appears to do well in its confrontations with pulsar
observations.  However, there is one family of recent observations which seems to present an
extremely embarassing problem for it, and we turn next to that question.

\section{Long Period Free Precesison in Some Radiopulsars?}

Year-period non-sinusoidal oscillations observed in some radiopulsar pulse shapes, $\dot
P$, and $P$ have been interpreted as NS free precession [Stairs, Lyne, \& Shemar 2000, Cordes
1993, Lyne et al. 2001] (or at least that of the conducting crust in which the surface
${\mb{B}}$-field is fixed) because it seemed the only plausible explanation.  But this
gives rise to a paradox: because of the interaction between the solid crust's
nuclei and its internuclear superfluid neutrons'  vortex lines, crustal precession
should have periods or damping times orders of magnitude smaller than a year
[Shaham 1977, Sedrakian, Wasserman, \& cordes 1999, Link \& Epstein 2000].  The 
strong interaction
 between the {\it core's} vortices and flux tubes, together with the crustal pinning of
core magnetic field when it passes through the crust, makes this already too short time scale
even smaller and more difficult to evade.  This is a crucial problem which will raise doubts
about canonical descriptions of NS structure until it is solved\footnote{With this in
mind {A possible but very different interpretation of this
``precession" may be that the observed radiopulse structure oscillations are not
caused by precession of the NS but only that of parts of its magnetospheric current
pattern which slowly rotates relative to the NS external magnetic field.} Such a
motion has already been suggested [Ruderman \& Sutherland 1975] as the origin of the well
known radiopulsar ``drifting sub-pulse'' phenomenon [Backer 1973, Desphande \& Rankin 1999]:
${\mb{v}} = {\mb{E}} \times {\mb {B}} c/B^2$ drift inside a radiopulsar's polar
cap (PC) accelerator gives the $e^\pm$ outflow above the accelerator and the
currents within it an additional spin ${\mb{\Omega}}_d \sim \Delta V
\Phi^{-1}\hat{\mb{B}} \sim 10 \hat{\mb{B}}{\rm s}^{-1}$. ($\Phi$ is the open
field line flux through the PC accelerator and $\Delta V (\sim 10^{12}V)$ is the
potential drop through it.)  With non-axisymmetric surface fields and
gravitaitonal bending of PC initiated $\gamma$-rays, or possibly $\gamma$-rays
produced from ``outergap'' accelerators, $e^\pm$ associated current flows can exist on open
${\mb{B}}$-field lines which do not pass through a PC accelerator before reaching
the NS surface.  These will, however, need a distributed $e\Delta V \sim mc^2$
along their flow to adjust the local net charge density to that needed to keep ${\mb{E}} \cdot
\hat{\mb{B}} \sim 0$. For such flows
$\Omega_d \sim mc^3/e\Phi \sim 2\pi{\rm yr}^{-1}$ near the needed ``precession'' rate.
(There remains the hard long stnading question of the deveopment of non-uniform current
distributions so that $\mb{\Omega}_d$ is reflected in modulations of $P$, $\dot P$, and the
shape of radioemisison pulses.)}.

\section{An Acknowledgement}

Some of the work summairzed here was carried on and presented during past visits to
Amsterdam. In all of many interactions there, and elsewhere, with Jan van Paradijs. Jan's
initiative and intelligence were always combined with great warmth and friendliness.  I am
extremely grateful for having known him.


\begin{references}



\reference Alpar,  M.A., Anderson, P., Pines, D., \& Shaham, J. 1984, ApJ, 282,
791 and Alpar, M., Chau, H.F., Cheng, K.S., \& Pines, D. 1993, ApJ, 409, 305.

\reference Backer, D., 1973, ApJ, 182, 245.

\reference Camilo, F., Kaspi, V., Lyne, A., Manchester, R., Bell, J., D'Amico, N., mcKay, N.,
\& Crawford, F. 2000, ApJ, 541, 367.

\reference Chen, K., Ruderman, M., \& Zhu, T. 1998, ApJ, 493, 397.

\reference Chen, K. \& Ruderman, M. 1993, ApJ, 408, 179.

\reference Cordes, J. 2000, ITP Neutron Star Workshop (Santa Barbara).

\reference Deshpande, A., \& Rankin, J. 1999, ApJ, 524, 1008.

\reference Ding, K., Cheng, K.S., \& Chau, H. 1993, ApJ, 408, 167.

\reference Jahan-Miri, M. 2000, ApJ, 532, 514.

\reference Jayawardhana, R., \& Grindlay, J.E. 1996,  Astron. Soc. Pac. Conf. Ser., 105, 231.

\reference Jones, P. 1998, MNRAS 296, 217.

\reference Kaspi, V.M., Manchester, R.N., Siegman, B., Johnston, S., \& Lyne, A.G.
1994, ApJ, 422, L83.

\reference Konenkov, D., \& Geppert, U. 2001, MNRAS, 325, 426.

\reference Link, B. and Cutler, C. 2001, MNRAS, in press.

\reference Link, B. \& Epstein, R. 2001, ApJ, 556, 392.

\reference Lundgren, S. 1995, ApJ, 453, 433.

\reference Lyne, A.G., Graham Smith, F., \& Pritchard, R.S. 1992, Nature, 369,
706.

\reference Lyne, A.G., Pritchard, R.S., \& Smith, F.G. 1993, MNRAS 265, 1003.


\reference Lyne, A.G., Pritchard, R.S., \& Smith, F.G. 1993, MNRAS 265, 1003.

\reference Lyne, A.G., Pritchard, R.S., Smith, F.G., \& Camilo, F. 1996, Nature,
381, 497.


\reference Lyne, A.G. et al. 2001, Nature, in press.

\reference McCulloch, P., Hamilton, P., McConnell, D., \& King, E. 1990, Nature,
346, 822.

\reference \"Ogelman, H., Finley, J.P., \& Zimmerman, H.U. 1993, Nature, 361, 
136.

\reference Ruderman, M. 1991, ApJ, 366, 261 and ApJ, 382, 576.

\reference Ruderman, M. \& Sutherland, P. 1975, ApJ, 196, 51.

\reference Ruderman, M., Zhu, T., and Chen, K. 1998, ApJ, 492, 267 (and erratum 1998).

\reference Sauls, J. 1989, in {\sl Timing Neutron Star}, eds. H. \"Ogelman and E.
van den Heuvel (Dordrecht: Kluwer).

\reference Sedrakian, A., Wasserman, I., and Cordes, J. 1999, ApJ, 524, 341.

\reference Shaham, J. 1977, ApJ, 214, 251.

\reference Srinivasan, G., Bhattacharya, D., Muslimov, A., \& Tsygan, A. 1990,
Current Sci., 59, 31.

\reference Stairs, I., Lyne, A., and Shemar, S. 2000, Nature, 406, 484.

\reference Trumper, J. 2001, these proceedngs.

\reference Van den Heuvel, E.P.J., \& Bitzaraki, O, 1995, A\&A, 297, L41.

\reference Zhang, W., Marshall, F., Gotthelf, E., Middleditch, J., \& Wang, Q. 2001, ApJ, 544,
L177.

\end{references}
\end{document}